# Performance-Aware Power Management in Embedded Controllers with Multiple-Voltage Processors

[1,4]Feng Xia, [2]Liping Liu, [3]Longhua Ma, [3]Youxian Sun and [1]Jinxiang Dong

[1]College of Computer Science and Technology, Zhejiang University, Hangzhou 310027, China
[2]School of Electrical Engineering and Automation, Tianjin University, Tianjin 300072, China
[3]State Key Lab of Industrial Control Technology, Zhejiang University, Hangzhou 310027, China
[4]Faculty of Information Technology, Queensland University of Technology, Brisbane QLD 4001, Australia

*Name of Corresponding Author:* Feng Xia
*Complete Postal Address:* College of Computer Science and Technology, Zhejiang University, Hangzhou 310027, China
*E-mail:* f.xia@ieee.org
*Alternate E-mail:* f.xia@acm.org

## Abstract

The goal of this work is to minimize the energy dissipation of embedded controllers without jeopardizing the quality of control (QoC). Taking advantage of the dynamic voltage scaling (DVS) technology, this paper develops a performance-aware power management scheme for embedded controllers with processors that allow multiple voltage levels. The periods of control tasks are adapted online with respect to the current QoC, thus facilitating additional energy reduction over standard DVS. To avoid the waste of CPU resources as a result of the discrete voltage levels, a resource reclaiming mechanism is employed to maximize the CPU utilization and also to improve the QoC. Simulations are conducted to evaluate the performance of the proposed scheme. Compared with the optimal standard DVS scheme, the proposed scheme is shown to be able to save remarkably more energy while maintaining comparable QoC.

## Introduction

Applications of embedded processors in various engineering systems have been expanding rapidly in recent years. With continuous miniaturization of physical size, an increasing number of embedded processors are battery powered. For these energy-limited systems, power management is important because low power dissipation not only prolongs the battery's lifetime but also increases the system reliability. Even if the energy constraint does not exist, energy should also be conserved to reduce the operational cost and environmental effect.

With the advent of mobile computing and wireless networking techniques, the use of battery-powered embedded processors in real-time control systems is rapidly growing. A typical example is mobile robots. Consequently, control engineers are confronted with a 'new' type of resource management problem, i.e., power management. In general, however, low energy consumption and high-quality system performance are conflicting with each other (Pillai and Shin, 2001). Therefore, in embedded controllers the energy consumption must be minimized in a way that the Quality-of-Control (QoC) of the system is not jeopardized.

Significant effort has been made on power and energy management in general-purpose embedded systems. The majority of the achieved results take advantage of the dynamic voltage scaling (DVS) technology (Aydin *et al.*, 2004; Gaujal and Navet, 2007). By adjusting the operating voltage and frequency of the processors dynamically, DVS has proved effective in energy conservation since the energy dissipation is approximately proportional to the square of the voltage (i.e., $E \propto V^2$). Regardless, limited work has been done in power management for real-time control applications. As one of the first work in this direction, Lee and Kim (2005) formulated the power management in multitasking control systems as an optimization problem, and proposed a static solution and a dynamic solution for the problem. Jin *et al.* (2007) presented a feedback fuzzy-DVS scheduling architecture integrating feedback control and fuzzy DVS for real-time control tasks. Recently, we have also developed several DVS algorithms for real-time control systems (Xia and Sun, 2006; Xia *et al.*, 2008).

A common assumption of the above-mentioned approaches is that the supply voltage of the processor could be varied continuously. However, this is not the case for modern processors: only a limited number of voltage levels are practically available. With these multiple-voltage processors, applying DVS algorithms that assume continuous voltage levels to embedded control systems cannot realize the full potential of energy reduction (Hua and Qu, 2005). To guarantee the system schedulability, a waste of computing resources may potentially result from the quantization of the voltage levels. Recently, Marinoni and Buttazzo (2007) proposed a method that combines discrete DVS management with elastic scheduling of control tasks to fully utilize the available CPU resources, but QoC-awareness is not considered in the method. In contrast, this paper focuses on exploiting performance-aware application adaptation to reduce energy consumption.

This paper deals with power management in embedded control systems running on processors with multiple voltage/speed levels. A QoC-aware power management scheme will be developed based on the DVS technique. To achieve further energy consumption reduction over standard DVS, the proposed scheme adapts the periods of control tasks according to current QoC of the corresponding control loops. The *ideal* voltage level is computed using DVS and the minimum among the voltage levels that are not lower than the ideal one is chosen for the processor. To avoid resource waste, a resource reclaiming mechanism is utilized, which decreases the task periods to maximize the CPU utilization and also to improve the QoC whenever possible. By exploiting application adaptation, the proposed approach can achieve significant additional energy reduction over standard DVS, while maintaining comparable QoC. The effect of discrete voltage levels will also be attacked. Simulation results will be given to validate the effectiveness of the proposed approach.

## System Model

Consider a DVS-enabled embedded controller, which is responsible for executing $N$ independent control tasks $\{\tau_i\}$ concurrently. Each control task corresponds to a physical process. The supply voltage and operating frequency, i.e. CPU speed, can be scaled with a scaling factor $\alpha \in \{\alpha_1, \alpha_2, ..., \alpha_M\}$, where $\alpha_M = 1$, and $\alpha_m < \alpha_{m+1}$. Hereafter, $\alpha$ will also be used to denote the (normalized) CPU speed. The timing attributes of $\tau_i$ are as follows.

- $h_i$: period, which equals the sampling period of the control loop $i$, with a nominal (initial) value of $h_{i,0}$.
- $c_{i,nom}$: nominal execution time at full CPU speed, i.e. when $\alpha = 1$.
- $c_i$: actual execution time when CPU speed is adapted, and it holds that $c_i = c_{i,nom}/\alpha$.

By default, the relative deadline of a control task is equal to its period. In this paper we use the terms *task period* and *sampling period* interchangeably since they are always equal. The CPU *utilization* $U = \sum c_i / h_i$, while the CPU *workload* $\omega = U \cdot \alpha = \sum c_{i,nom} / h_i$. The system utilizes the Earliest Deadline First (EDF) algorithm as the underlying scheduling policy. Accordingly, the schedulability condition is:

$$\sum_{i=1}^{N} \frac{c_i}{h_i} \leq 1 \Leftrightarrow \omega \leq \alpha \quad (1)$$

To guarantee the feasibility of solutions, assume that $\sum_{i=1}^{N} \frac{c_{i,nom}}{h_{i,0}} \leq 1$. The switching overheads between different voltage levels of the processor is neglected, since in most cases the switching time of prevailing processors is negligibly small in comparison with task periods. The (normalized) energy consumption of the processor is calculated as (Sinha and Chandrakasan, 2001):

$$E(\alpha) = \alpha^2 \quad (2)$$

## Performance-Aware Power Management

In this section, we will devise a performance-aware power manager for the above-described system. The power manager (Fig. 1) consists of three modules: period adaptation, voltage scaling, and resource reclaiming. The first module is responsible for adjusting the sampling period of each loop based on feedback information about its current QoC. This module is activated every time a new *job* (or *task instance*) of any control task is released. Whenever the workload level changes due to variations in the period and/or nominal execution time of any task, the second module decides a new speed for the processor using DVS. If the CPU is not fully utilized, the last module will then re-adjust the task periods to reclaim unused CPU resources.

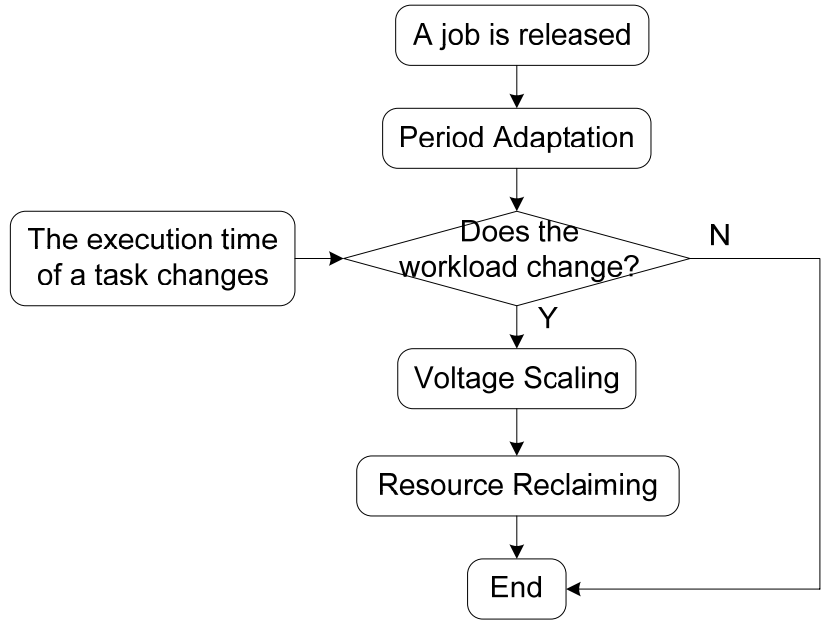

Fig. 1: Flow chart of performance-aware power management

**Period Adaptation:**

Increasing task periods will reduce the CPU workload level. Under the schedulability constraint of the task set, i.e. (1), it is possible to choose a lower speed level for a lower workload, thus achieving additional energy reduction. Actually, it has been shown that increasing task periods benefits energy savings (Xia and Sun, 2006). According to sampled-data control theory, however, larger sampling periods yield worse QoC with the traditional framework of fixed timing constraints. Therefore, the effectiveness of saving energy by increasing sampling periods is often impaired by significant QoC deterioration.

Previous studies have shown that in many control systems, when the control errors are relatively small (i.e. the QoC is good), the sampling periods may be enlarged to some extent without significant QoC degradation (Buttazzo *et al.*, 2007). From this insight, we propose to adapt task periods to QoC, while aiming to reduce the energy consumption as much as possible. The sampling period of a control loop will be properly increased whenever the QoC is good enough.

Every time a new job of $\tau_i$ is released, the corresponding task period will be adjusted as follows:

$$h_i(k) = \eta_i(k) \cdot h_{i,0} \quad (3)$$

where $k$ denotes the sampling instant of control loop $i$, and $\eta_i$ is the *period scaling factor* given by:

$$\eta_i(k) = \begin{cases} h_{i,\max}/h_{i,0} & \text{if } e_i(k) \leq e_{i,\min} \\ \dfrac{e^{-\beta \cdot e_i(k)} - e^{-\beta \cdot e_{i,\max}}}{e^{-\beta \cdot e_{i,\min}} - e^{-\beta \cdot e_{i,\max}}}(h_{i,\max}/h_{i,0} - 1) + 1 & \text{if } e_{i,\min} < e_i(k) < e_{i,\max} \\ 1 & \text{if } e_i(k) \geq e_{i,\max} \end{cases} \quad (4)$$

where $e_i(k)$ is *absolute control error* of control loop $i$ at the $k$-th instant, which is defined as the absolute difference between the reference input $r_i(k)$ and the system output $y_i(k)$, i.e., $e_i(k) = |r_i(k) - y_i(k)|$; $h_{i,max}$ is the maximum allowable sampling period of control loop $i$, which could be determined e.g. using well-established control theory on system stability or simply by simulations; $\beta$ is a constant introduced to enhance the effect of the exponential function; $e_{i,min}$ and $e_{i,max}$ are design parameters that decide the range of $e_i$ in which the sampling period $h_i$ can be adjusted arbitrarily.

Once $e_i(k)$ exceeds the upper threshold $e_{i,max}$, the sampling period will be directly set to the minimum, which corresponds to $\eta_i(k) = 1$. This enables improving control performance via quick responses to large derivations. In contrast, the period will be the maximum, i.e. $\eta_i(k) = h_{i,max}/h_{i,0}$, if $e_i(k)$ becomes less than the limit $e_{i,min}$, implying that the system approaches a steady state. The maximum period is set to achieve the largest possible energy saving. In other cases, a period that decreases exponentially with increasing $e_i(k)$ will be assigned.

As $\beta$ increases, the effect of the exponential function will be enhanced. Since a small $\eta$ value yields a relatively small sampling period, the algorithm becomes more aggressive with smaller $\beta$, from the viewpoint of energy saving. The principle of choosing $e_{min}$ and $e_{max}$ is to use as large values as possible given that the QoC is not jeopardized.

**Voltage Scaling:**

The voltage scaling module will be invoked if only the workload changes, which may result from the intentional period adaptation and/or unintentional execution time variations. Upon every invocation, this module first computes the ideal voltage level and then chooses a speed available from the discrete range $\{\alpha_1, \alpha_2, ..., \alpha_M\}$.

The ideal CPU speed is calculated by:

$$\alpha_{ideal} = \sum_{i=1}^{N} \frac{c_{i,nom}}{h_i} = \omega \quad (5)$$

With a standard DVS scheme assuming continuous voltage levels, it is the best way to scale voltage according to (5). Eq. (5) yields the maximum CPU utilization (i.e. 100%) and lowest possible energy expenditure while preserving the system schedulability (Sinha and Chandrakasan, 2001).

After the ideal CPU speed is computed, the CPU speed will then be set to:

$$\alpha = \begin{cases} \alpha_1 & \text{if } \alpha_{ideal} < \alpha_1 \\ \alpha_m & \text{if } \alpha_{ideal} = \alpha_m \\ \alpha_{m+1} & \text{if } \alpha_m < \alpha_{ideal} < \alpha_{m+1} \end{cases} \quad (6)$$

**Resource Reclaiming:**

The discreteness of available voltage levels may cause actual CPU resources to be underutilized. This is primarily because, when $\alpha_m < \alpha_{ideal} < \alpha_{m+1}$, it is imperative to set CPU speed to $\alpha_{m+1}$ in order not to violate task schedulability. As a consequence, the expected CPU utilization will be $U_{exp} = \alpha_{ideal}/\alpha_{m+1} < 100\%$. In addition, the minimum allowable CPU speed $\alpha_1$ may cause unused computing resources as well.

In this paper, a simple resource reclaiming mechanism is employed to avoid significant resource waste particularly when the number of voltage levels is relatively small. The basic idea behind this mechanism is to properly decrease task periods when the CPU utilization is expected to be less than 100% so that the free CPU resources are also exploited. In this module, the sampling periods are updated as follows:

$$\tilde{h}_i = \frac{\alpha_{ideal}}{\alpha} h_i \quad (7)$$

It can be seen that (7) attempts to rescale all task periods with $\alpha_{ideal}/\alpha$. With this resource reclaiming mechanism, the resulting CPU utilization will then be:

$$U = \sum \frac{c_i}{\tilde{h}_i} = \sum \frac{c_{i,nom}/\alpha}{h_i \cdot \alpha_{ideal}/\alpha} = \frac{1}{\alpha_{ideal}} \sum \frac{c_{i,nom}}{h_i} = \frac{\omega}{\alpha_{ideal}} = 100\% \quad (8)$$

Consequently, the CPU will be fully utilized. With the resource reclaiming module, the waste of resources due to discrete voltage levels is avoided, and the QoC of each loop can be improved through decreasing sampling periods.

## Performance Evaluation

This section conducts simulation experiments using Matlab/TrueTime (Cervin *et al.*, 2003) to evaluate the performance of the proposed scheme. The system simulated has four independent control loops/tasks, i.e. $N = 4$. The controllers are of PID (Proportional-Integral-Derivative) type, with parameters $K_P$, $K_I$ and $K_D$ (Xia *et al.*, 2007). The system models of the controlled processes and the corresponding controller parameters are given in Table 1. The following parameters are used in all simulation experiments: $\beta = 40$, $e_{max} = 0.3$, $e_{min} = 0.02$. To measure the QoC quantitatively, the Integral of Absolute Error (IAE) is recorded respectively for each loop, i.e., $J_i(t) = \int_0^t e_i(\tau)d\tau$. The total control cost of the system is calculated as $J_{SUM}(t) = \sum_{i=1}^{4} J_i(t)$. Normalized CPU energy consumption is computed using (2).

In the simulations, all loops are perturbed by step input changes at the same time. The perturbation interval is set to 1s. The whole simulation lasts 12s every time. Consider five different processors that support the following normalized voltage levels:

- CPU-1: {0.5, 1.0}
- CPU-2: {0.45, 0.64, 0.92, 1.0} (Hua and Qu, 2005)
- CPU-3: {0.36, 0.55, 0.64, 0.73, 0.82, 0.91, 1.0} (Pillai and Shin, 2001)

- CPU-4: {0.285, 0.333, 0.380, 0.428, 0.476, 0.523, 0.571, 0.619, 0.666, 0.714, 0.761, 0.809, 0.857, 0.904, 0.952, 1.0} (Soria-Lopez *et al.*, 2005)
- CPU-*ideal*: An ideal processor in which the voltage can be scaled arbitrarily within the range of (0, 1].

Table 1: Settings of the simulated control system

| Loop No. | System Model | Controller Parameters | $c_{nom}$ (ms) | $h_0$ (ms) | $h_{max}$ (ms) |
|---|---|---|---|---|---|
| 1 | $\frac{1}{1000s+50}$ | $K_P=10^4$, $K_I=400$, $K_D=0$ | 2 | 10 | 40 |
| 2 | $\frac{1}{s^2+10s+20}$ | $K_P=30$, $K_I=70$, $K_D=0$ | 2 | 7 | 30 |
| 3 | $\frac{1}{0.5s^2+6s+10}$ | $K_P=100$, $K_I=200$, $K_D=2$ | 2 | 8 | 30 |
| 4 | $\frac{1}{s^2+10s+20}$ | $K_P=200$, $K_I=350$, $K_D=3$ | 2 | 9 | 40 |

For the purpose of comparison, the simulation results of the *optimal* standard DVS scheme (denoted osDVS) for CPU-*ideal* will also be given. With this scheme, the task periods will be fixed without online adaptation. The ideal CPU speed will be set using (5). As mentioned previously, this scheme is optimal among all standard DVS schemes.

**Results and Analysis:**

Fig. 2 depicts the energy consumption of different processors. The optimal standard DVS scheme results in a constant (normalized) energy consumption of 0.918 due to fixed workload. When the proposed approach is used, the energy consumption varies at runtime. Compared with osDVS, all processors using our approach consume much less energy most of the time. Due to the close relationship between the energy consumption and the CPU speed (i.e. $E = \alpha^2$), the variations in the CPU speed can also be observed from Fig. 2.

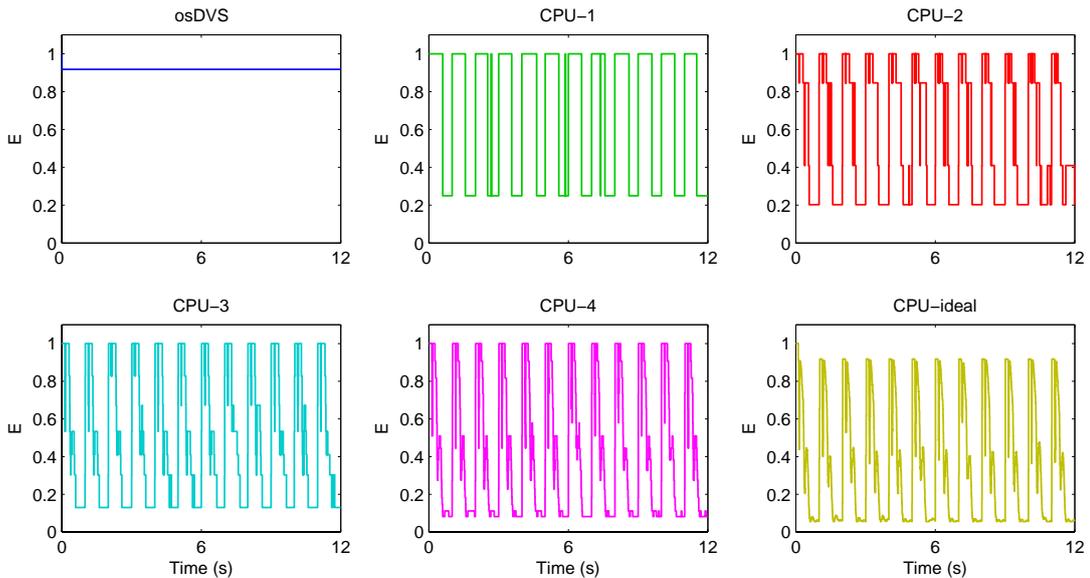

Fig. 2: Normalized energy consumption

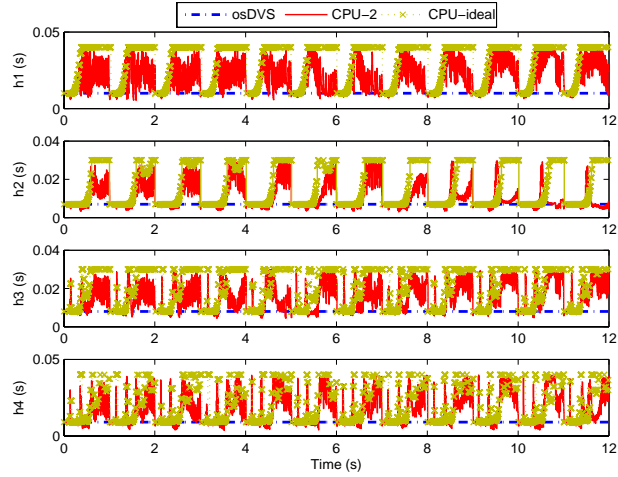

Fig. 3: Periods of control tasks

In contrast to osDVS that uses fixed task periods, the proposed scheme dynamically adjusts the task periods (Fig. 3), which results in varying energy consumption (Fig. 2). In most situations, our approach exploits larger task periods than osDVS. As a general rule, the task periods with CPU-ideal are the largest among all the studied cases. This explains why CPU-ideal consumes the least energy (see also Table 2).

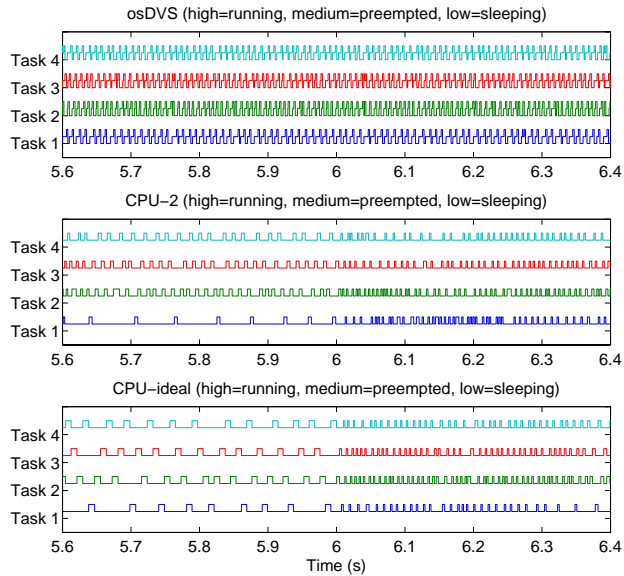

Fig. 4: Schedule of different processors

As shown in Fig. 4, the CPU schedule also demonstrates the changes in task periods. It can be seen that all task periods (i.e. the time interval between two consecutive activations) are fixed under osDVS, whereas the task periods vary over time when our scheme is employed. Furthermore, larger periods correspond to larger execution times that imply lower CPU speeds and energy consumption, and vice versa.

The difference in the QoC of each loop (Table 1) caused by the use of different processors with different schemes is fairly minor. For example, the IAE of loop 1 in different cases is 1.205, 1.211, 1.243, 1.242, 1.252, and 1.298, respectively. When our scheme is used in the four processors with multiple voltages, the maximum increase in the IAE of loop 1 relative to osDVS is 3.9%. The average energy consumption $E_{AVG}$ and the total control cost $J_{SUM}$ are summarized in Table 2. With the ideal processors, our scheme achieves an additional average energy reduction of 0.414 relative to osDVS (the relative reduction is 45.1%), while the total control cost increases only by 7.6%. Even with the CPU-1 that allows only two voltage levels, our scheme still consumes 13.3% less energy than the osDVS with an ideal processor, and the relative increase in total control cost is only 4.1%.

Table 2: Average energy consumption and total control cost

| Performance Index | osDVS | QoC-aware power management |       |       |       |           |
|-------------------|-------|-------|-------|-------|-------|-----------|
|                   |       | CPU-1 | CPU-2 | CPU-3 | CPU-4 | CPU-*ideal* |
| $E_{AVG}$         | 0.918 | 0.796 | 0.694 | 0.636 | 0.614 | 0.504     |
| $J_{SUM}$         | 7.588 | 7.895 | 8.034 | 8.006 | 8.020 | 8.164     |

The effect of the number of available voltage levels can be observed from Table 2. As the number of available voltage levels increases, the performance of the proposed scheme in energy saving becomes better. For instance, the proposed scheme yields the least energy dissipation for the ideal processor CPU-*ideal*; the CPU-3 with 7 voltage levels consumes an average energy 20.1% less than that of CPU-1 with 2 voltage levels. The reduction in energy consumption is achieved with a small penalty in QoC. For instance, the total control cost with CPU-3 increases only 1.4% in comparison with the case of CPU-1.

It can be summarised from the above simulation results that: 1) the proposed performance-aware power management scheme can achieve significant additional energy reduction over the optimal standard DVS scheme, while delivering comparable control performance; 2) the proposed scheme is applicable to various processors supporting different voltage levels and more voltage levels are beneficial to energy savings.

## Conclusion

A performance-aware power management scheme has been developed for real-time embedded control systems with multiple-level voltage processors. From the runtime QoC of each control loop, the scheme exploits period adaptation to achieve additional energy reduction over standard DVS. To make full use of available CPU resources, a resource reclaiming mechanism has been employed. Simulation results have shown that the proposed scheme is cost-effective in energy saving for multiple-voltage embedded controllers.

## Acknowledgement

The first author would like to thank Prof Yu-Chu Tian at QUT, Australia, for helpful discussions. This work is partially supported by China Postdoctoral Science Foundation under Grant No. 20070420232, Natural Science Foundation of China under Grant No. 60474064, Zhejiang Provincial Natural Science Foundation of China under Grant No. Y107476, and Australian Research Council under Discovery Projects Grant No. DP0559111.

## References

Aydin, H., R. Melhem, D. Mosse and P. Mejia-Alvarez, 2004. Power-Aware Scheduling for Periodic Real-Time Tasks. IEEE Trans. Comput., 53(5): 584-600

Buttazzo, G., M. Velasco and P. Marti, 2007. Quality-of-Control Management in Overloaded Real-Time Systems. IEEE Trans. Comput., 56(2): 253-266

Cervin, A., D. Henriksson, B. Lincoln, J. Eker and K.-E. Arzen, 2003. How Does Control Timing Affect Performance. IEEE Control Syst. Mag., 23(3): 16-30.

Gaujal, B. and N. Navet, 2007. Dynamic voltage scaling under EDF revisited. Real-Time Syst., 37: 77-97

Hua, S. and G. Qu, 2005. Voltage Setup Problem for Embedded Systems with Multiple Voltages. IEEE Trans. Very Large Scale Integr. (VLSI) Syst., 13(7): 869-872.

Jin, H., D.L. Wang, H.A. Wang and H. Wang, 2007. Feedback fuzzy-DVS scheduling of control tasks. J. Supercomput., 41(2): 147-162

Lee, H.S. and B.K. Kim, 2005. Dynamic Voltage Scaling for Digital Control System Implementation. Real-Time Syst., 29: 263-280.

Marinoni, M. and G. Buttazzo, 2007. Elastic DVS Management in Processors with Discrete Voltage/Frequency Modes. IEEE Trans. Ind. Inform., 3(1):51-62.

Pillai, P. and K.G. Shin, 2001. Real-Time Dynamic Voltage Scaling for Low Power Embedded Operating Systems. In Proc. 18th Symp. Operating System Principles, ACM, Banff, Alberta, Canada, 21-24 October 2001, pp. 89-102.

Sinha, A. and A.P. Chandrakasan, 2001. Energy efficient real-time scheduling. In Proc. Int. Conf. Computer Aided Design, IEEE/ACM, San Jose, California, USA, 4-8 Nov. 2001, pp. 458-463,

Soria-Lopez, A., P. Mejia-Alvarez and J. Cornejo, 2005. Feedback Scheduling of Power-Aware Soft Real-Time Tasks. In Proc. 6th Mexican Int. Conf. Computer Science (ENC'05), IEEE CS, Puebla, Mexico, 26-30 Sept. 2005, pp. 266-273.

Xia, F., G.S. Tian and Y.X. Sun, 2007. Feedback Scheduling: An Event-Driven Paradigm. ACM Sigplan Not., 42(12): 7-14.

Xia, F., Y.-C. Tian, Y.X. Sun and J.X. Dong, 2008. Control-Theoretic Dynamic Voltage Scaling for Embedded Controllers. IET Comput. Digit. Tech., in press. DOI:10.1049/iet-cdt:20070112.

Xia, F. and Y.X. Sun, 2006. An Enhanced Dynamic Voltage Scaling Scheme for Energy-Efficient Embedded Real-Time Control Systems. Lect. Notes Comput. Sci., 3983: 539-548.